\documentclass{raa}
\usepackage{graphicx, times, amsmath, natbib}
\def\aj{{AJ}}                   
\def\apj{{ApJ}}                 
\def\apjl{{ApJ}}                
\def\apjs{{ApJS}}               
\def\ao{{Appl.~Opt.}}           
\def\aap{{A\&A}}                
\def\aaps{{A\&AS}}              
\def\mnras{{MNRAS}}             
\def\prd{{Phys.~Rev.~D}}        

\def\jcap{{JCAP}} 

\def\hmpc{\, h^{-1}{\rm Mpc}}

\begin{document}
\title{On the fairness of the main galaxy sample of SDSS}

\volnopage{Vol.0 (201x) No.0, 000--000}
\setcounter{page}{1}

\author{Kelai Meng\inst{1} \and Bin Ma\inst{2} \and Jun Pan\inst{1}
\and Longlong Feng\inst{1}}
\institute{The Purple Mountain Observatory, 2 West Beijing Road,
  Nanjing 210008, China\\
  \and
  The National Astronomical Observatories, 20A Datun Road, Beijing 100012, China}

\date{Received~~2010 month day; accepted~~2010~~month day}

\abstract{
Flux-limited and volume-limited galaxy samples are constructed from
SDSS data releases DR4, DR6 and DR7 for statistical analysis.
The two-point correlation functions $\xi(s)$, monopole of three-point
correlation functions $\zeta_0$, projected two-point correlation function
$w_p$ and pairwise velocity dispersion $\sigma_{12}$ are measured to test
if galaxy samples are fair for these statistics.
We find that with increment of sky coverage of SDSS,
$\xi(s)$ of flux-limited sample is extremely robust and
insensitive to local structures at low redshift. But
for volume-limited samples fainter than $L^*$ at large scales $s>\sim 10\hmpc$,
deviation of $\xi(s)$ and $\zeta_0$ of DR7 to those
of DR4 and DR6 increases with larger absolute magnitude. In the weakly
nonlinear regime, there is no agreement between $\zeta_0$ of
different data releases in all luminosity bins. Furthermore,
$w_p$ of volume-limited samples of DR7 in luminosity bins fainter than
$-M_{r,0.1}=[18.5,19.5]$  are significantly larger, and
$\sigma_{12}$ of the two faintest volume-limited samples of DR7 display
very different scale dependence than results of DR4 and DR6.
Our findings call for cautions in understanding clustering analysis results
of SDSS faint galaxy samples, and higher order statistics of SDSS volume-limited
samples in the weakly nonlinear regime. The
first zero-crossing points of $\xi(s)$ of volume-limited samples are also investigated
and discussed.
}

\authorrunning{Meng et al.}
\titlerunning{On the fairness of the main galaxy sample of SDSS}

\maketitle

\section{Introduction}
Clustering analysis of galaxy samples thrives for the availability of modern
massive galaxy surveys. The two mostly successful and biggest galaxy surveys
to date are the two-degree field galaxy redshift survey
\citep[2dFGRS, ][]{CollessEtal2003} and the Sloan Digital Sky
Survey \citep[SDSS, ][]{YorkEtal2000}. The final data
release of the 2dFGRS offers 3-D mapping of roughly a quarter of million galaxies,
the SDSS achieves spectra of $\sim 0.9$ million galaxies \citep{AbazajianEtal2009}.
The unprecedented number of galaxies and enormous volume surveyed by SDSS defines its
unique role in the era of precision cosmology \citep{KomatsuEtal2010}, by its power
spectra and the two-point correlation functions (2PCF) at large scales
\citep[e.g.][]{TegmarkEtal2004b, EisensteinEtal2005, PercivalEtal2010, ReidEtal2010}.

Another highly appreciated application of clustering analysis of galaxies
is to relate galaxy distribution to dark matter and halos, aiming at
inferring processes galaxies experienced during their formation and
evolution. Interpretation of statistics of galaxy samples provided by SDSS
prevails in category of the $\Lambda$CDM+halo model and relevant extensions
such as the halo occupation distribution
\citep[HOD, e.g.][]{BerlindWeinberg2002, KravtsovEtal2004, ZhengEtal2005}
and the conditional luminosity function
\citep[CLF, ][]{YangEtal2003}. For example, works of
\citet{ZehaviEtal2002, ZehaviEtal2005, ZehaviEtal2010} systematically explored the
luminosity
and color dependence of galaxy 2PCFs and extensively quantified HOD parameters of
galaxies; \citet{Cooray2006} derived the
occupation of central and satellite galaxies in halos and their corresponding
conditional luminosity functions from
a compilation of correlation functions of SDSS, attempting to draw clues of
galaxy evolution with reference to high redshift samples;
\citet{LiEtal2007} rather directly compared projected correlation functions and the
pairwise velocity dispersion (PVD) of
SDSS with those of mock galaxy samples populated from N-body simulations by
semi-analytic models (SAM) of \citet{KangEtal2005} and
\citet{CrotonEtal2006}, they find that SAM can roughly reproduced observed
clustering of SDSS galaxies but have to reduce faint satellite fraction in massive
halos in the prescription of SAM by $\sim 30$ percent to resolve discrepancies in PVD.

Yet there are challenges to the fairness of SDSS galaxy samples, i.e. whether
galaxy samples of SDSS are complete and have enough volume to be a fair
representation of the Universe.
In fact prudence in reading out physics from measured statistics
especially correlation functions has been called, \citet{NicholEtal2006} disclosed that exclusion
of the {\em Sloan Great Wall} \citep[at $z\sim 0.08$, ][]{GottEtal2005} would
change the 2PCF by $\sim 40\%$ and the three-point correlation function
(3PCF) by as much as $\sim 70\%$ of the sample defined by the {\it r}-band
absolute magnitudes $-22\leq M_{r, 0.1} \leq -19$. The apparent influence of
super structures on estimated correlation functions at large scales is
somehow against intuition since one already takes it for granted that
the SDSS galaxy sample's depth and sky coverage is sufficient to accomplish
homogeneity, spatial averaging
would suppress the variance induced by a particular structure in a small patch of
sky. \citet{SylosEtal2009} noticed that the zero-crossing point of 2PCF
of SDSS main galaxy sample varies with luminosity and sample depth, and
anti-correlation is absent in the mostly recent measured 2PCF of SDSS
luminous red galaxy (LRG) sample
\citep[e.g.][]{MartinezEtal2009, KazinEtal2010}.
By the extreme-value statistical analysis, \citet{AntalEtal2009}
purport that either the SDSS suffers from severe
sample volume dependent intrinsic systematical effects or there is persistent density
fluctuation not fading away over scales beyond standard $\Lambda$CDM model
prediction.

It is therefore important for one to check the fairness of galaxy samples used
in order to endorse the confidence of relevant analysis.
It is understood that fairness means differently for
different statistics and also samples constructed in various ways.
\citet{ZehaviEtal2010} have laboriously
evaluated finite volume effects and impact of super structures, they
compared 2PCFs of volume-limited galaxy sub-samples in full depth
with of the same sub-sample but limited in a smaller volume overlapped with
the volume-limited sub-sample defined in the luminosity bin one dex lower. Their
experiment leads to the conclusion that finite volume effects are insignificant for
anisotropic and projected 2PCFs in nonlinear regime for their
sub-samples of luminosity higher than $M_{r,0.1} = -19$. But they then
find that including faint galaxies causes weird regulation to 2PCF, which is
similar but of smaller amplitude to the discovery in
\citet{ZehaviEtal2005} using an early release of SDSS. We notice that
such analysis for galaxies of luminosity lower than $-18$ is missed though 2PCF of their
faintest sub-sample $M_{r,0.1}\in[-18, -17]$ is
adopted for estimation of biasing and HOD parameters.

These works mainly concentrate on changes to two-point statistics
by altering sample depth, we rather check the fairness by sky coverage enlargement,
not only of 2PCFs but also of monopole of 3PCFs in redshift space, projected 2PCFs and
PVDs. There are data releases 4, 6 and 7 of SDSS main galaxy catalogue
\citep[DR4, DR6 and DR7 by][respectively]{AdelmanEtal2006,
AdelmanEtal2008, AbazajianEtal2009},
the increment of sky coverage from DR4 to DR6 is roughly the same from DR6 to DR7.
An advantage to investigate effects of sample volume on correlation function
by sky angular coverage other than survey depth is that the restriction of apparent
magnitudes of the survey definition limits permitted range of depth adjustment, in
particular for those faint galaxies which are visible only at low redshift and
support very shallow sample space. And one of our purpose is to see how correlation
function evolves {\em naturally} with the progress of a real survey.

Section 2 describes SDSS data and estimation
methods of statistics  we used, results are shown in section 3, the last section
is of summary and discussion.

\section{Galaxy Samples and Estimation of Correlation Functions}
\subsection{sample construction}

\begin{figure*}
\resizebox{\hsize}{!}{\includegraphics{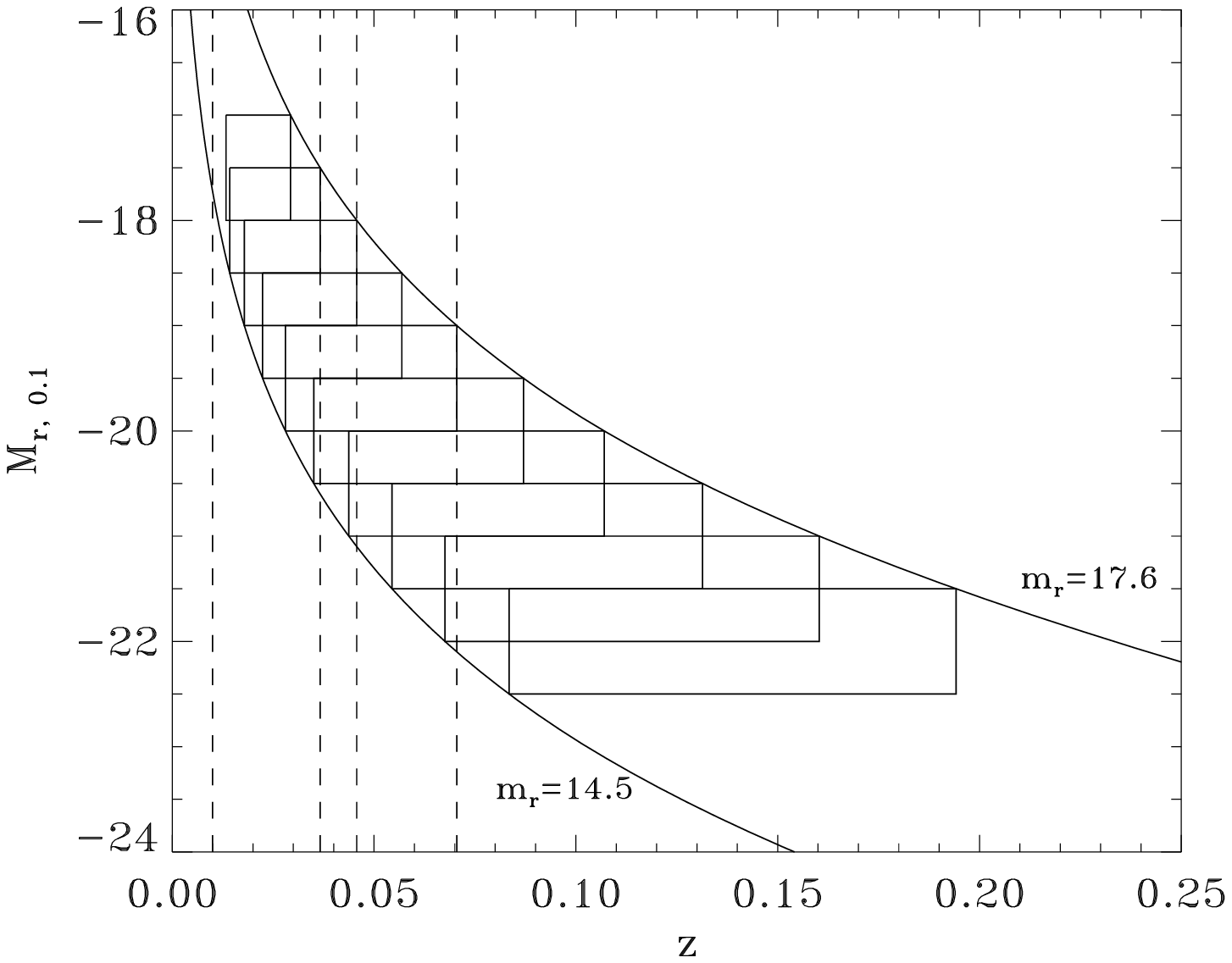}
\includegraphics{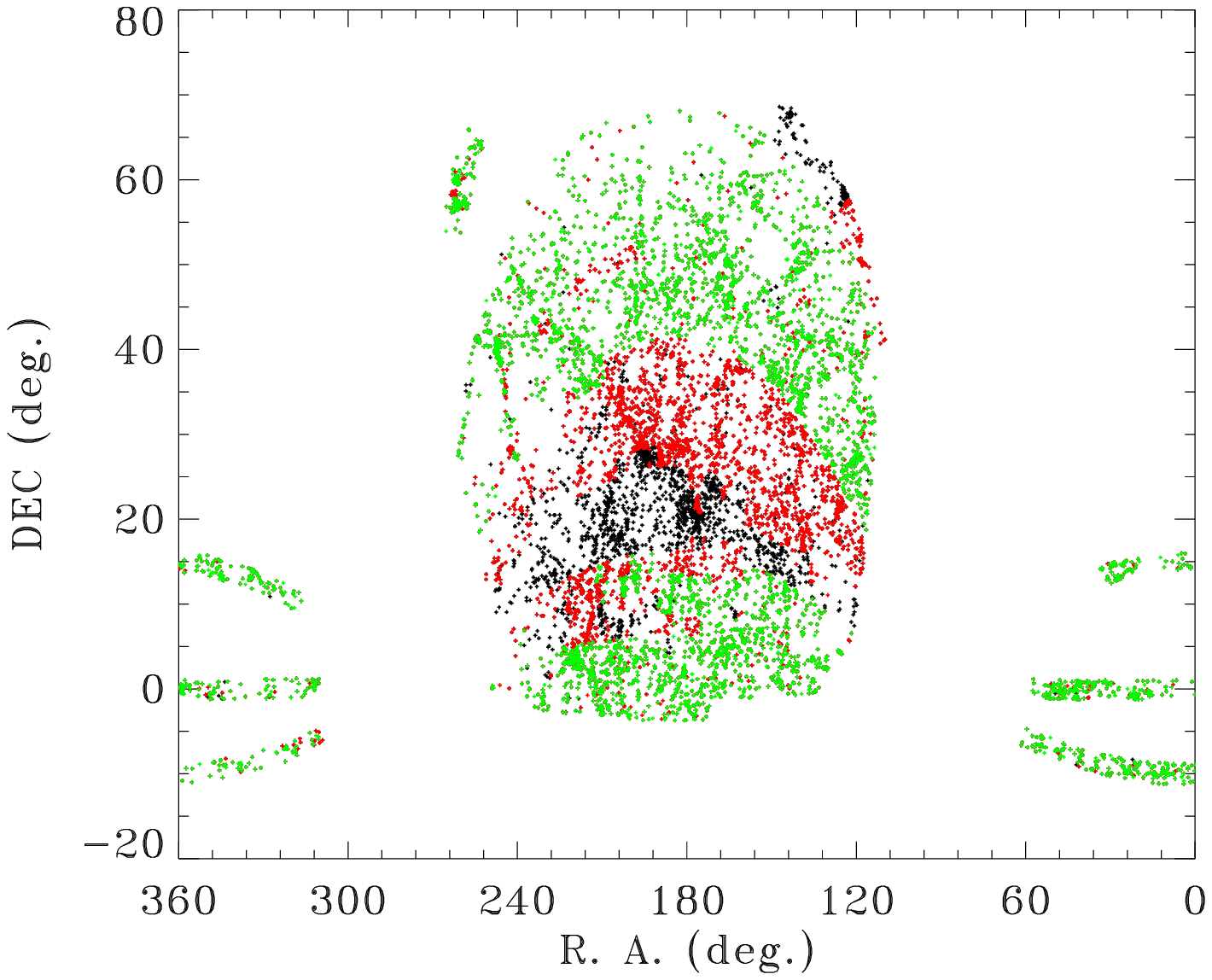}}
\caption{Left panel shows the definition of SDSS galaxy subsamples
on the redshift-absolute magnitude plane,  the two curves are
boundaries of VAGC catalogue resulted from the apparent
magnitude limits imposed, those overlapping
rectangles delineate where volume-limited samples located, and dashed lines
label the lower redshift cuts of our flux-limited samples.
In the right panel, distributions of galaxies of volume-limited subsample
$-M_{r,0.1}=[17,18]$ on celestial sphere are plotted, green points are galaxies of
SDSS DR4, red points indicate extra galaxies in DR6 and black point are galaxies
added in DR7.}
\label{fig:defsample}
\end{figure*}

The {\tt safe} galaxy sample of the New York University Value-Added Galaxy
Catalog \citep[NYU-VAGC,][]{BlantonEtal2005} \footnote{{\tt
http://sdss.physics.nyu.edu/vagc}} is a catalog of low redshift
galaxies (mostly below $z \sim 0.3$) defined by apparent magnitudes of
$14.5<m_r<17.6$. Three data releases
in chronological order are selected, namely DR4, DR6 and DR7, which
spectroscopic surveyed areas are about $4,783$, $6,860$ and
$8,032$ square degrees respectively. As spectroscopic coverage of SDSS is
not uniform, we use only those regions of spectroscopic completeness
greater than $0.9$. We did not perform fibre collision correction
to improve completeness, the correction only becomes significant at
scales $< 0.2h^{-1}$Mpc for SDSS galaxies \citep{ZehaviEtal2002}.
To ensure the correct geometry, galaxies
in the three catalogues are also filtered with their own accompanied survey windows,
bright star masks and completeness masks.

Flux-limited samples defined by {\it r}-band apparent magnitude
range $14.5 < m_r < 17.6$ and redshift $0.01 < z <0.23$ are generated.
Consequently we obtain $300,661$ galaxies in DR4, $447,407$ in DR6,
and $535,845$ in DR7. In order to explore influence of
local galaxies on correlation functions, we
also constructed confined flux-limited galaxy samples by
near-end redshift cut of $z_{min} = 0.037, 0.046, 0.071$.
Volume-limited sub-samples are also produced in consecutive luminosity
bins starting from $M_{r,0.1}=-17$ to
$-22.5$ in step of 0.5 magnitude and bin width of one magnitude, the
absolute magnitude in NYU-VAGC is corrected to redshift $z = 0.1$
and is $K$ corrected, but e-correction is not taken into account.
We noticed that there are some galaxies having different apparent magnitudes in DR7
than in early data releases, so we constrcuted a couple of
additional volume-limited samples from DR7 but filtered with masks of
DR4 for comparison, measurements indicate that
such differences have little influence on statistics employed.

Details of these samples are shown in
Tables~\ref{tab:fl} \& \ref{tab:vl} and Figure~\ref{fig:defsample},
comoving distances of galaxies are calculated in a flat $\Lambda$CDM
universe with $\Omega_m=0.3$, $\Omega_\Lambda=0.7$ and $h=0.7$.

\begin{table}
\caption{Numbers of galaxies in flux limited samples defined
by {\em r}-band apparent magnitude
$14.5 < m_r < 17.6$ and redshift $ z_{min}\leq z \leq0.23$.}
\begin{center}
\begin{tabular}{ccccc}
\hline
 $z_{min}$ & 0.010 & 0.037 & 0.046 & 0.071 \\
\hline
  DR4 & 300,661 & 281,400 & 268,247 & 216,373 \\
  DR6 & 447,407 & 417,426 & 397,543 & 321,915 \\
  DR7 & 535,845 & 498,445 & 473,980 & 382,921 \\
\hline
\end{tabular}
\end{center}
\label{tab:fl}
\end{table}

\begin{table}
\caption{Volume limited samples. Distances are in in unit of $\hmpc$.}
\begin{center}
\begin{tabular}{lcccccccccc}
\hline
Label & Luminosity  &\multicolumn{2}{c}{redshift} & & \multicolumn{2}{c}{comoving distance}
 & & \multicolumn{3}{c}{number of galaxies}\\
\cline{3-4}\cline{6-7}\cline{9-11}
 & $M_{r,0.1}-5\log_{10}h$ & $z_{min}$ & $z_{max}$ & & $d_{min}$ & $d_{max}$ &
&DR4 & DR6 & DR7 \\
\hline
VL1 &  $[-18.0,-17.0]$ & 0.011 & 0.029 & & 33.89 & 87.31 & &4,223 & 6,389 & 8,219 \\
VL1+&   $[-18.5,-17.5]$ & 0.014 & 0.037 & & 42.53 & 108.95 & & 7,292 & 11,543 & 14,343 \\
VL2 &  $[-19.0,-18.0]$ & 0.018 & 0.046 & & 53.32 & 135.61 & & 11,639 & 18,328 & 22,500 \\
VL2+&  $[-19.5,-18.5]$ & 0.022 & 0.057 & & 66.77 & 168.27 & &19,209 & 29,463 & 35,932 \\
VL3 &  $[-20.0,-19.0]$ & 0.028 & 0.071 & & 83.51 & 207.96 & &31,807 & 47,565 & 57,363 \\
VL3+&  $[-20.5,-19.5]$ & 0.035 & 0.087 & & 104.24 & 255.70 & &50,719 & 75,162 & 89,654 \\
VL4 &  $[-21.0,-20.0]$ & 0.044 & 0.107 & & 129.83 & 312.96 & &59,215 & 87,295 & 103,924 \\
VL4+&  $[-21.5,-20.5]$ & 0.054 & 0.131 & & 161.21 & 381.91 & &60,132 & 89,602 & 107,207 \\
VL5 &  $[-22.0,-21.0]$ & 0.068 & 0.160 & & 199.41 & 462.71 & &46,264 & 69,499 & 82,239 \\
VL5+&  $[-22.5,-21.5]$ & 0.083 & 0.194 & & 245.46 & 555.88 & &24,002 & 36,677 & 43,631 \\
\hline
\end{tabular}
\end{center}
\label{tab:vl}
\end{table}

\subsection{Estimation of correlation functions}
\subsubsection{Redshift space correlation functions}

Isotropic 2PCF $\xi(s)$ of separation $s$ in redshift
space is measured with the estimator of \citet{LandySzalay1993},
\begin{equation}
\xi=\frac{DD-2DR+RR}{RR}\ , \label{eq:LS}
\end{equation}
in which $DD$, $RR$ and $DR$ are respectively the normalised numbers
of weighted galaxy-galaxy, random-random and galaxy-random pairs at
given separation. To proceed the estimation with Eq.~\ref{eq:LS},
corresponding random sample is generated following
distributions of redshift, magnitude, geometric constraints, spectroscopic
completeness and survey masks of each individual galaxy sample but with 20 times
of numbers of points. Each galaxy and random point is assigned a weight
according to their redshift and angular position to minimize the
variance in estimated $\xi$ \citep{Efstathiou1988, Hamilton1993},
\begin{equation}
w_i=\frac{1}{1+4\pi\,n(z)\Phi_i\,J_3(s)}\ ,
\end{equation}
where $\Phi_i$ is the selection function at the location of $i$th
galaxy, $n(z)$ is the mean number density, and
$J_3(s)=\int_{0}^{s}\xi(s)s^2ds$. The $J_3(s)$ is computed using a
power-law $\xi(s)$ with correlation length $s_0=8\hmpc$ and
$\gamma_0=1.2$\citep{ZehaviEtal2002}.

Calculation of 3PCFs of all those galaxy samples
lasts too long, we turn to measure the monopole of the 3PCF instead
\citep{PanSzapudi2005b}, which is a degenerated version of 3PCF defined as
\begin{equation}
\zeta_0(s_1, s_2)=2\pi \int_{-1}^1 \zeta(s_1, s_2, \theta){\rm d}\cos\theta \ ,
\end{equation}
and estimated via
\begin{equation}
\zeta_0=\frac{DDD-3DDR+3DRR-RRR}{RRR}\ ,
\label{eq:ss}
\end{equation}
where combined symbols of D and R are normalized numbers of triplets counted
within and between data sets of galaxies and random points, e.g. if the number
of galaxies around galaxy $i$
in bin $(s_1^{lo}, s_1^{hi})$ is $n_i(s_1)$,
in bin $(s_2^{lo}, s_2^{hi})$ is $n_i(s_2)$,
the $DDD$ in Eq.~\ref{eq:ss} reads
\begin{equation}
DDD=\left\{ \begin{array}{cc} \frac{\sum_{i=1}^{N_g} n_i(s_1) n_i(s_2)}{N_g (N_g
-1) (N_g-2)}, & if \, s_1 \ne s_2 \\
\frac{\sum_{i=1}^{N_g} n_i(s_1) \left(n_i(s_2)-1\right)}{N_g (N_g-1) (N_g-2)}\ ,
 & if \, s_1 = s_2 \end{array} \right.\ .
\end{equation}

\subsubsection{Projected 2PCF and PVD}
To minimize the effect of redshift distortion due to galaxy's peculiar
motion, the separation $s$ (or $r$ in real space) is divided into two
components, the parallel
part $\pi$ and the perpendicular part $\sigma$ with respect to line-of-sight,
the anisotropic 2PCF is measured on grids of $(\sigma, \pi)$.
Integration of $\xi(\sigma, \pi)$ over $\pi$ then yields a redshift
distortion free function, the projected 2PCF,
\begin{equation}
w_p(\sigma)=\int_{-\pi_{max}}^{+\pi_{max}}\xi(\sigma,\pi)d\pi
=\sum_i\xi(\sigma,\pi_i)\Delta{\pi_i}\ ,
\end{equation}
which has practically an integration limit $\pi_{max}=50\hmpc$.

It is well known that the redshift distortion consists of two components
dominated in different regimes, coherent infall is responsible for
the clustering enhancement at large scales while the smearing of
correlation strength at small scales is attributed to random
motions. At large scales the boost to the 2PCF by the peculiar velocities
takes a particularly simple form \citep{Kaiser1987, Hamilton1992},
\begin{equation}
\xi^\prime(\sigma,\pi)=\xi_0(s)P_0(\mu)+\xi_2(s)P_2(\mu)+\xi_4(s)P_4(\mu)\
, \label{eq:infall}
\end{equation}
where $P_\ell(\mu)$ is Legendre polynomials, $\mu$ is the cosine
of the angle between $r$ and $\pi$. Assuming $\xi
=(r/r_0)^{-\gamma}$ there are relations
\begin{equation}
\begin{aligned}
\xi_0(s)& =\xi(s)=\left(1+\frac{2\beta}{3}+\frac{\beta^2}{5}\right)\xi(r) \\
\xi_2(s)& =\left(\frac{4\beta}{3}+\frac{4\beta^2}{7}\right)
\left(\frac{\gamma}{\gamma-3}\right)\xi(r) \\
\xi_4(s)& =\frac{8\beta^2}{35}
\left(\frac{\gamma(2+\gamma)}{(3-\gamma)(5-\gamma)}\right)\xi(r)\ ,
\end{aligned}
\label{eq:kaiser}
\end{equation}
where $\beta\approx\Omega_0^{0.6}/b$, and $b$ is the linear bias
parameter, note that the first equation is independent on the functional
form of $\xi(r)$.

To incorporate effects of random motion, the anisotropic 2PCF in
redshift space is approximated by a convolution of
$\xi^\prime(\sigma,\pi)$ in Eq.~\ref{eq:infall} with the
distribution function of the pairwise velocity $f(v_{12})$
\citep[c.f.][]{Peebles1993},
\begin{equation}
\xi(\sigma,\pi)=\int_{-\infty}^{+\infty}
\xi^\prime(\sigma,\pi-\frac{v_{12}}{H_0})f(v_{12})dv_{12}\ ,
\label{eq:fog}
\end{equation}
and in general $f(v_{12})$ is assumed to obey an exponential
distribution with PVD  $\sigma_{12}$
\begin{equation}
f(v_{12})=\frac{1}{\sigma_{12}\sqrt{2}}\exp \left(-\frac{\sqrt{2}
v_{12}}{\sigma_{12}}\right)\ . \label{eq:fv}
\end{equation}

The parameter $\beta$ usually is derived from the ratio
of $\xi(s)$ to $\xi(r)$ at large scales via the first equation in
Eq.~\ref{eq:kaiser}, then other model parameters can be
determined by combining Eq.~\ref{eq:infall} -- Eq.~\ref{eq:fv} to
fit the $\xi(\sigma,\pi)$ data grids. Note that
\citet{JingEtal1998} assumed a slightly different exponential distribution
function of pairwise velocity which is followed by \citet{LiEtal2007}.

\subsubsection{Covariance Matrix}
Covariance matrices our results are
computed with the jack-knife technique \citep{Lupton1993, ZehaviEtal2002}. Each
galaxy sample is divided into twenty separate slices of approximately equal sky area,
then we perform the analysis twenty times, at each time leave a
different slice out. Covariance matrices are generated accordingly with these
twenty measurements, for instance, covariance of 2PCF
measured in two bins of $i$ and $j$ is simply
\begin{equation}
{\rm Cov}(\xi_i,\xi_j) =\frac{N-1}{N}\sum_{\ell=1}^{N}
(\xi_{i,\ell}-\overline{\xi_i})(\xi_{j,\ell}-\overline{\xi_j})\ ,
\end{equation}
in which $N=20$ is the number of jack-knife sub-samples we used.

\section{Results}

\subsection{Flux-limited samples}

\begin{figure}
\begin{center}
\resizebox{\hsize}{!}{\includegraphics{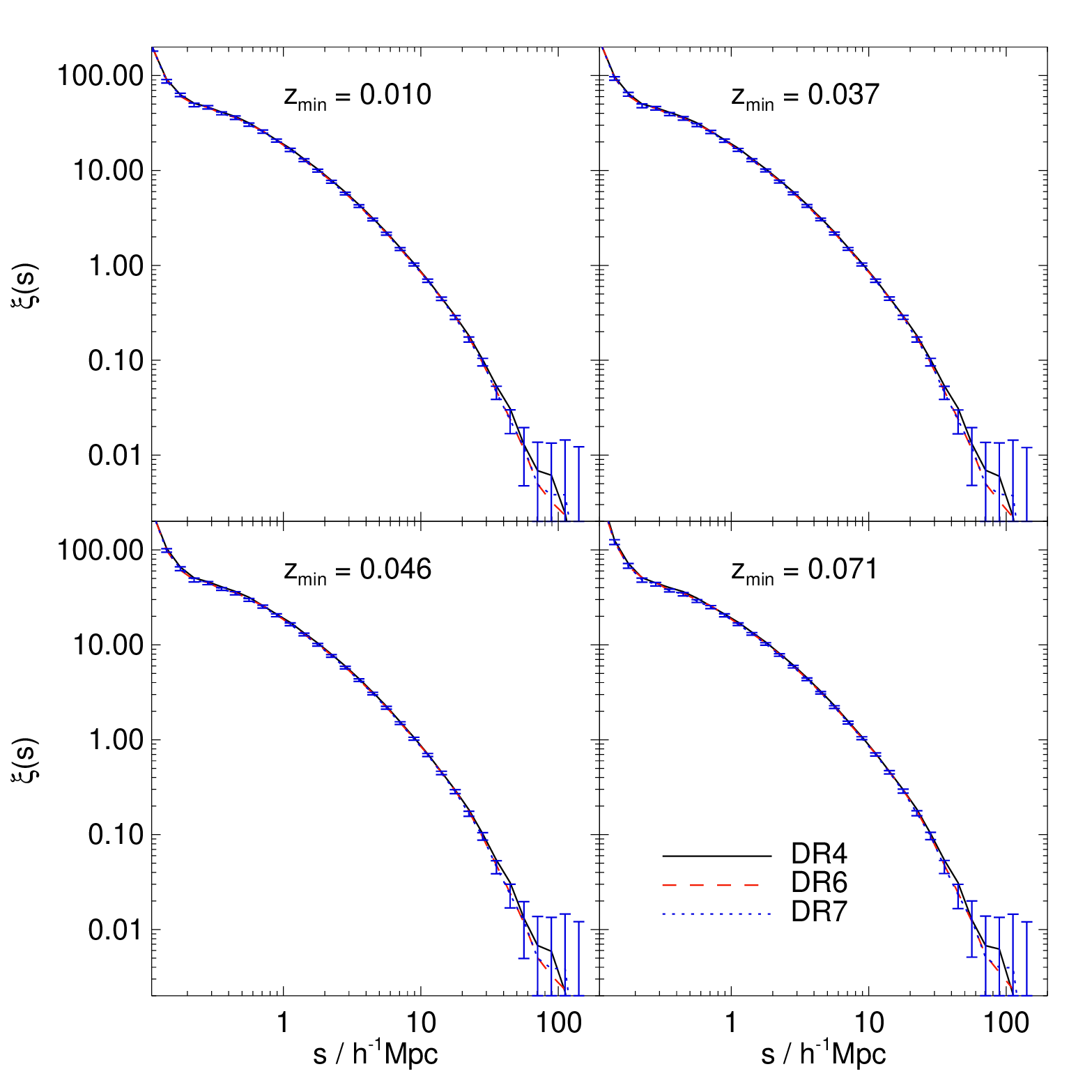}\includegraphics{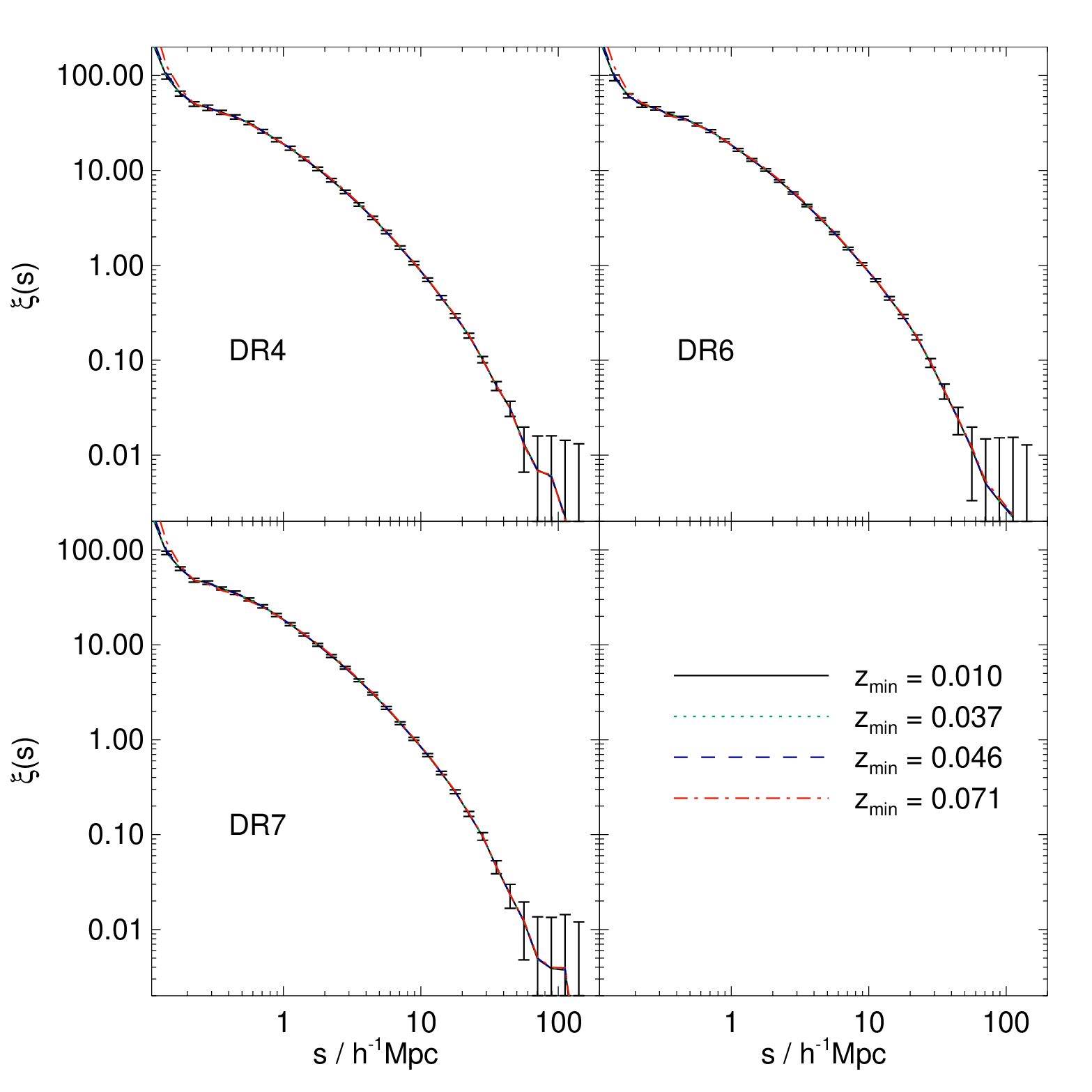}}
\end{center}
\caption{Redshift space 2PCFs of flux-limited samples.}
\label{fig:xis}
\end{figure}

Isotropic 2PCFs of flux-limited samples in Table~\ref{tab:fl}
are calculated firstly.
Figure~\ref{fig:xis} manifest that the redshift space 2PCFs of
flux-limited samples show little variation against data versions of SDSS.
$\xi(s)$ of DR4 exhibits some deviation at large scales $\sim 100 \hmpc$,
but is hardly significant for the huge cosmic variance at these scales.
$\xi(s)$ of flux-limited samples of the same data release are displayed
in the right panel of Figure~\ref{fig:xis}, there is no visible
change to redshift space 2PCF of SDSS when galaxies with low redshift are
excluded even when the lost of number of galaxies is as much as
$\sim 25\%$ (Table~\ref{tab:fl}).
Thus eliminating local volume and
enlarging sky coverage
from DR4 to DR7 have little influence on the clustering strength
measured, unlikely there is any significant sample volume
dependent effects. As we are not interested in general discussion
of the SDSS main galaxy catalogue as a whole, we
stop preforming further analysis with other statistical measures.

It is well known that faint galaxies have much lower linear bias
than luminous ones \citep[e.g.][]{TegmarkEtal2004b, ZehaviEtal2005, LiEtal2007},
when we throw away many faint galaxies by imposing near-end redshifts limit it is
expected that $\xi(s)$ should display higher amplitude when
lower redshift cuts increase. It could be that the lost in number of galaxies
(after proper weighting) is too small
to raise any serious deviation (Table~\ref{tab:fl}), or in another words $\xi(s)$
of flux-limited sample is dominated by galaxies around the redshift distribution peak.

\subsection{Volume-limited samples}
\subsubsection{2PCF and monopole of 3PCF in redshift space}
\begin{figure}
\resizebox{\hsize}{!}{\includegraphics{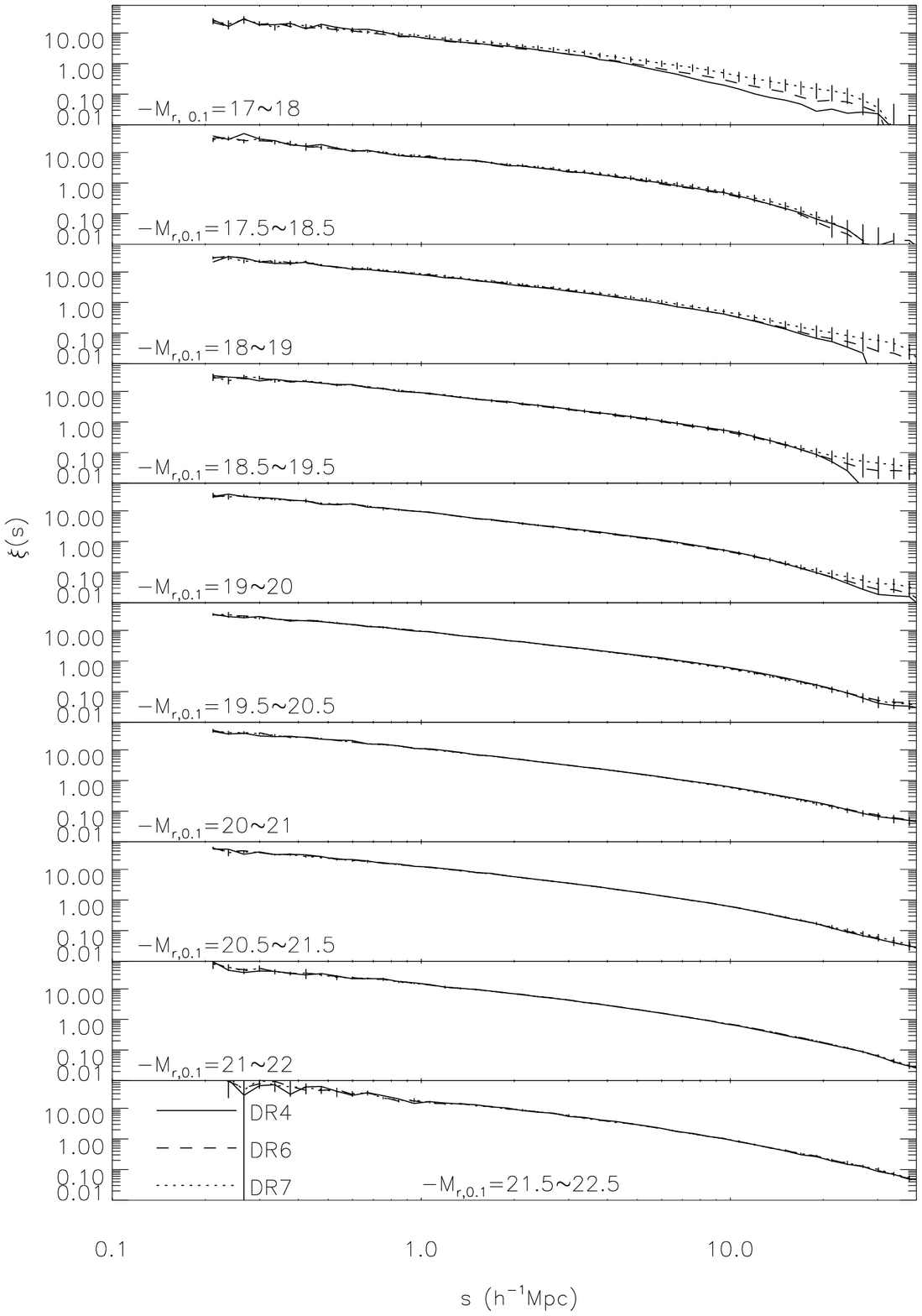}\includegraphics{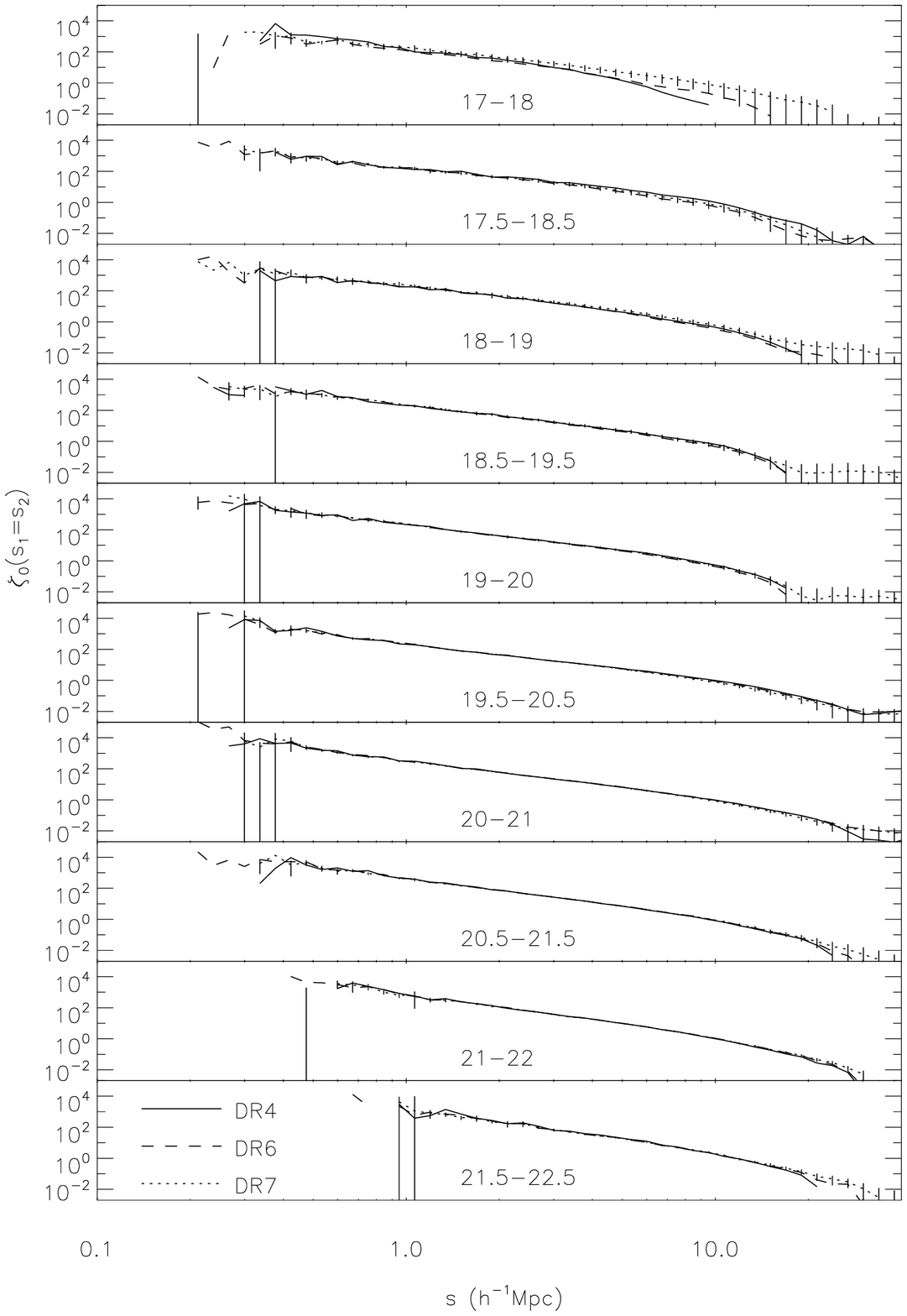}}
\caption{2PCFs $\xi(s)$ and monopoles of 3PCF $\zeta_0(s_1=s_2)$ at small scales
in redshift space of volume-limited samples.}
\label{fig:vlxinl}
\end{figure}

\begin{figure}
\resizebox{\hsize}{!}{\includegraphics{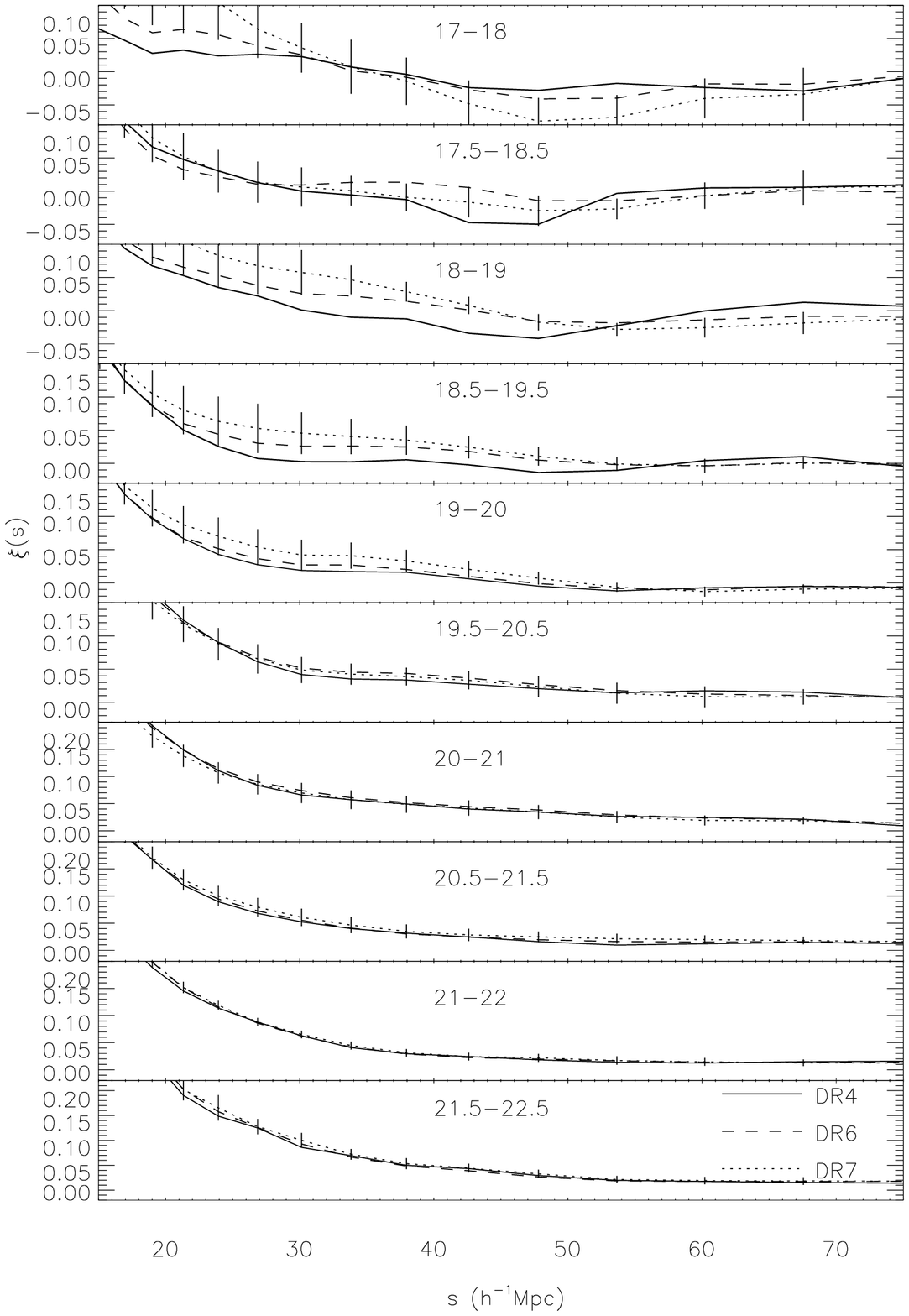}\includegraphics{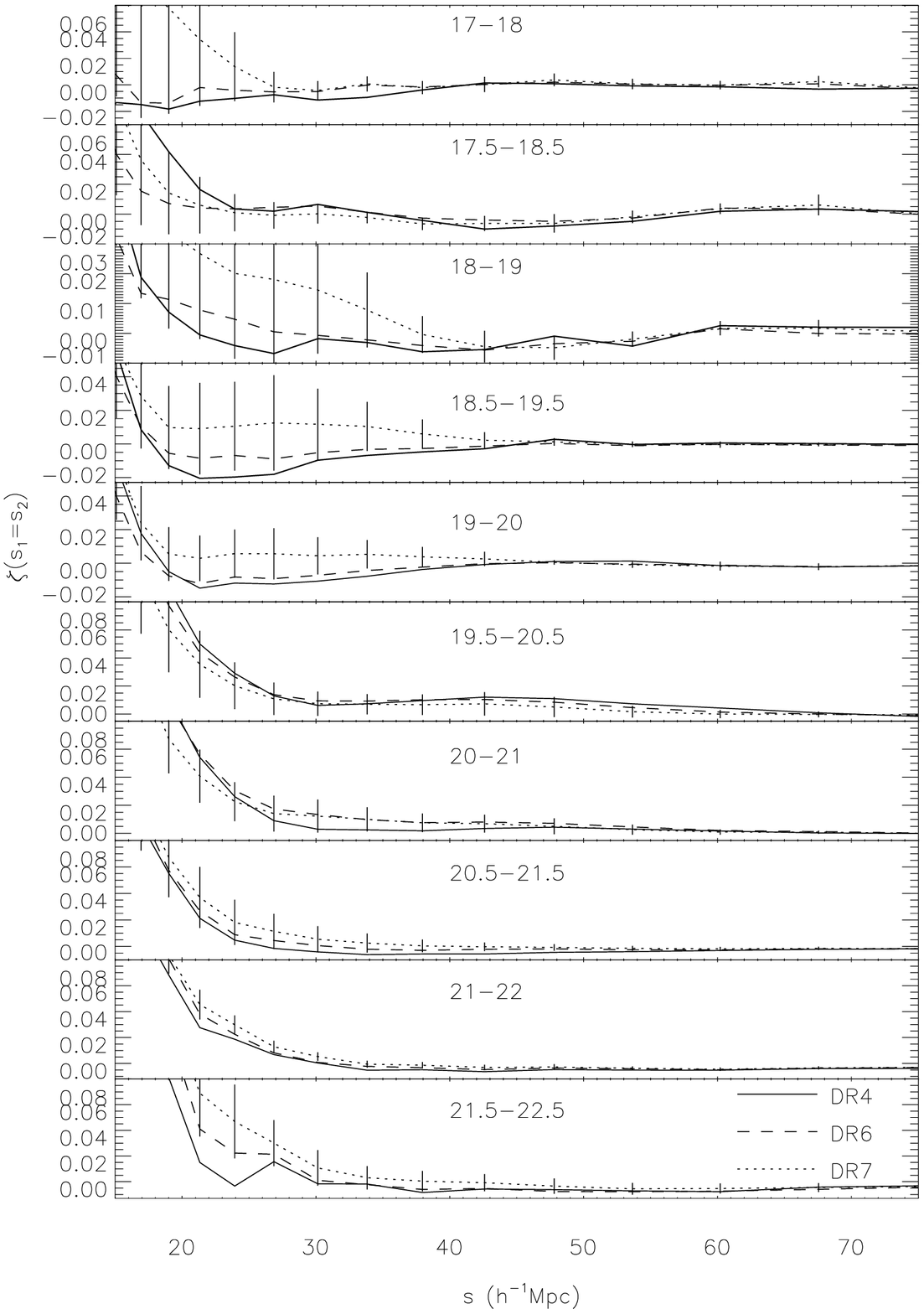}}
\caption{2PCFs $\xi(s)$ and monopoles of 3PCF $\zeta_0(s_1=s_2)$ at large scales
in redshift space of volume-limited samples.}
\label{fig:vlxilin}
\end{figure}

2PCFs $\xi(s)$ and monopoles of 3PCFs $\zeta_0(s_1,s_2)$ of volume-limited samples
of the three SDSS data releases are measured to probe possible differences. In this
paper we only present the $\zeta_0(s_1=s_2)$ which amplitude is
the strongest among configurations of $(s_1, s_2)$ \citep{PanSzapudi2005b}.
As seen in Figure~\ref{fig:vlxinl}, in nonlinear regime major discrepancies
appear in the VL1 sample of the lowest luminosity, differences between results of
DR4 and DR7 are around $2\sigma$ at scales as small as $\sim 3h^{-1}$Mpc, while
consistency of $\xi(s)$ and $\zeta_0$ of brighter volume-limited samples
of SDSS is perfect at scales $s<10\hmpc$.

At scales greater than $10\hmpc$, for subsamples of VL3+ -- VL5+
$\xi(s)$ of different data releases are in good agreement within errorbars, but
$\zeta_0$ have variations at level of $\sim 1\sigma$ (Figure~\ref{fig:vlxilin}).
For the five faint galaxy samples of VL1 -- VL3, disagreement in $\xi(s)$ of DR7 to
DR4 is already apparent in the regime, which is confirmed by their $\zeta_0$.
We conclude that modulation to correlation functions in redshift space resulted from
enlargement of sky coverage mainly occurs at scales ranging roughly from
$\sim 10$ to $\sim 50\hmpc$ where is usually classified as weakly
nonlinear regime in structure formation theory. Those applications and
conclusion appear somehow suspicious based on 3PCFs of volume-limited samples of
SDSS at large scales. For three-point correlation functions in redshift space,
fairness of volume-limited samples is guaranteed only at small scales, i.e. in
strongly nonlinear regime.

\subsubsection{The First Zero-crossing points of 2PCFs}

\begin{figure}
\begin{center}
\resizebox{0.75\hsize}{0.5\hsize}{\includegraphics{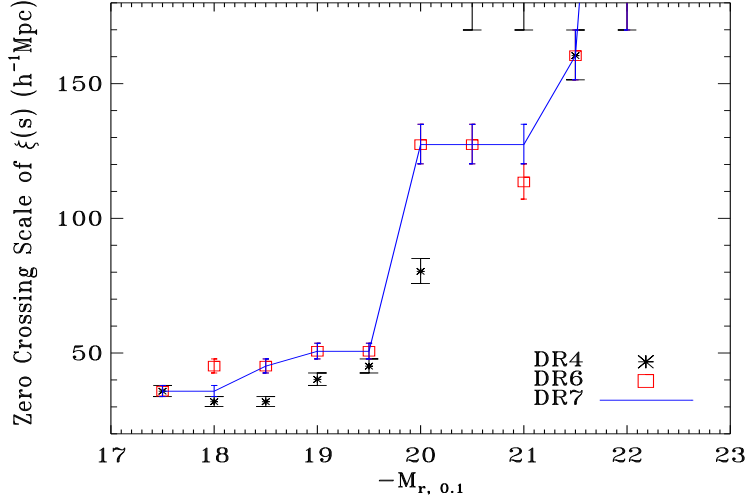}}
\end{center}
\caption{Luminosity dependence of the first zero crossing scales of $\xi(s)$ of
volume-limited samples. Lower caps of error bars are scales where $\xi>0$ and
higher caps of error bars are the adjacent scales where $\xi$ immediately becomes
negative, those points showing with only lower caps denote that the first zero crossing point
is actually larger than the scale probed in this work, larger than $\sim 170\hmpc$.}
\label{fig:vlxi0x}
\end{figure}

To investigate the charge of \citet{SylosEtal2009} the first zero crossing scales of
$\xi(s)$ against median luminosity of volume-limited samples are plotted in
Figure~\ref{fig:vlxi0x}. Estimated $\xi(s)$ is
effectively averaged over a scale bin $[s^{lo}, s^{hi}]$ and the quoted scale
is set to be $s=\sqrt{s^{lo}s^{hi}}$. Unlikely we can right hit all zero points
of $\xi(s)$ by our scale binning, so we choose to show
the range of scales within which $\xi(s)$ experiences zero-crossing which is drawn
as errorbars over the geometric mean of the pair of scales. From Figure~\ref{fig:vlxi0x}
it is clear that in general the brighter is
the characteristic luminosity of the sample, the larger the first zero crossing
scale will be. The five faint volume-limited samples
(VL1 -- VL3) have roughly the same first zero crossing scale
with mild variation between $\sim 30-50\hmpc$, then the crossing scale ascends abruptly
to as large as more than $100\hmpc$ and even higher than the largest scale we
measured ($\sim 170\hmpc$).

For faint galaxy samples, their depths are typically small and so be their effective
volumes, the systematical effect of integral constraint can not be
ignored \citep{LandySzalay1993, Bernstein1994}. In the weak correlation limit, the cosmic
bias resulted from integral constraint can be approximated by
\begin{equation}
b_{\xi}= \frac{\hat{\xi}}{\xi}-1
\approx -\frac{\bar{\xi}(R)}{\xi}, \ \ if\ |\xi|,\ |\bar{\xi}(R)|\ll 1\ ,
\end{equation}
in which $\hat{\xi}$ is the estimated 2PCF, $R$ is the smallest size of the sample and
$\bar{\xi}(R)$ is the average of the 2PCF over the sample volume, i.e. density variance at
sample volume \citep{LandySzalay1993}.
There is no {\it a priori} correction method to this bias unless we assume
something to model the shape of the 2PCF. Since $\bar{\xi}$ is
positive, naturally $\hat{\xi} \approx \xi-\bar{\xi}(R)$ will have a smaller
first zero-crossing scale than $\xi$. If as usual we assume that galaxy bias $b$
is linear and scale independent, $\hat{\xi}=b^2(\xi-\bar{\xi}(R))$, the correction to
the first zero-crossing scale depends on sample volume only.
As $\bar{\xi}$ decreases with scale slowly,
it is expected that the first zero-crossing scales of faint galaxy samples
will gradually become larger when sample volume increases, which is true for VL2 -- VL3.
However, surprisingly, it is not for VL1 and VL1+. The faintest two subsamples have the
smallest sample volume, but the first-zero crossing scale of VL1 does not change when SDSS
marching from DR4 to DR7, while of VL+ the scale of DR7 becomes smaller than of DR4.
Furthermore, the difference between depths of VL3 and VL3+ is not very
large (Table~\ref{tab:vl}), but the first zero-crossing scales of their $\xi$ differ hugely.
Integral constraint alone could not explain the findings.

The increment of sky coverage from DR4 to DR6 is approximately the same as the
gain from DR6 to DR7, the first zero crossing scales of DR7 only differ from DR6 slightly
in two luminosity bins, while DR4 does not agree with other data releases significantly,
which makes it difficult to clutch at other simple geometric explanation, such as assuming
fractal galaxy distribution. Ergodicity bias could not be resorted too,
for low luminosity samples with low characteristic redshift,
the correction $\Delta\xi$ is positive \citep{PanZhang2010} and
would push zero point to larger scales, which obviously contradicts observation.
Neither could be redshift distortion,
as on large scales redshift distortion acts on galaxy 2PCF as a multiplication.

The sudden change of the first zero-crossing scale from faint galaxies to bright galaxies
probably implies that the composition of faint galaxy samples
is very different compared with bright galaxy samples, which may be attributed to the
shifting of leading role from satellite galaxies to central galaxies
in samples brighter than $-M_{r,0.1}>20$ \citep{LiEtal2007}. Whatever the physical
mechanism is, mathematically the effect to 2PCF is fully packed into a simple function,
the galaxy bias.  The linear biasing model assumes that
on large scales the galaxy 2PCF $\xi_g=b\xi_m$ in which $b\neq 0$ is
a deterministic, scale independent bias parameter and $\xi_m$ is the 2PCF of dark matter,
obviously if the model holds, the zero point of $\xi_g$ will not change no matter
what $b$ could be, e.g. scale dependent.
If we presume that the problem of zero crossing is in biasing, then either stochastic
or nonlinear bias has to be invoked. Simple calculation indicates that
if we adopt the parametrization to bias of \citep{FryGaztanaga1993} and
include the second-order bias parameter in 2PCF, to the leading order the effect
is again multiplicative and can not shift the first zero point of 2PCF. It appears that
stochastic biasing have to be considered. Details of the calculation however
is beyond scope of this paper and will be presented elsewhere.

Another interesting aspect is that the first zero-crossing scales 2PCFs
of samples VL4 and VL4+ of DR4 are larger than the largest scale of our measurements,
but not of DR6 and DR7. The lack of anti-correlation in the two luminosity bins of DR4
is probably an evidence of the modulation due to the Sloan Great Wall as revealed by
\citet{ZehaviEtal2005} and \citet{ NicholEtal2006}, the increased
sky coverage of DR6 and DR7 just successfully weakens the influence of
the super structure \citep{ZehaviEtal2010}.

\subsubsection{Projected 2PCF and PVD}

\begin{figure}
\begin{center}
\resizebox{0.75\hsize}{!}{\includegraphics{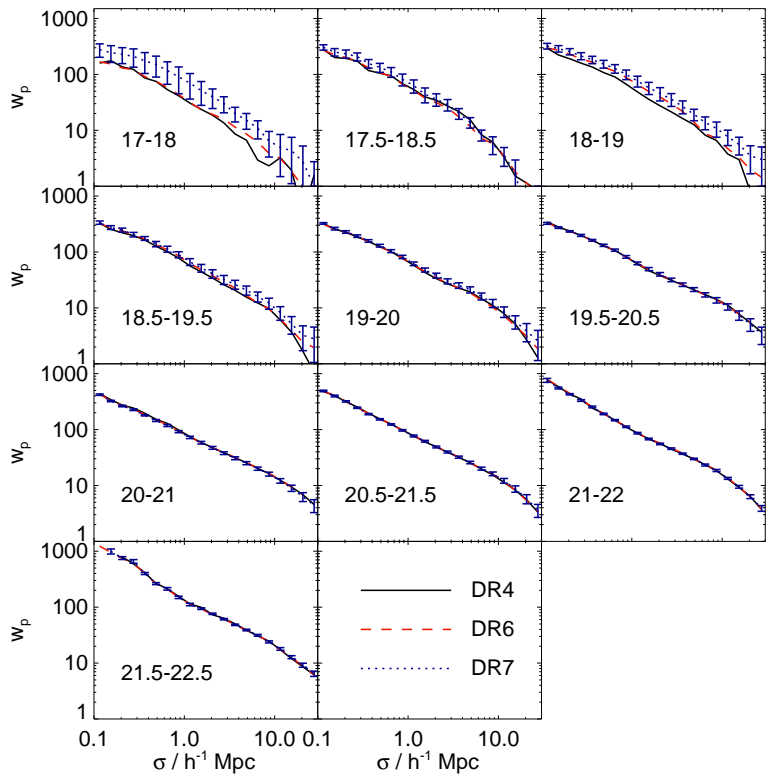}}
\end{center}
\caption{Projected 2PCFs of the volume-limited samples.}
\label{fig:wp}
\end{figure}

\begin{figure}
\begin{center}
\resizebox{0.75\hsize}{!}{\includegraphics{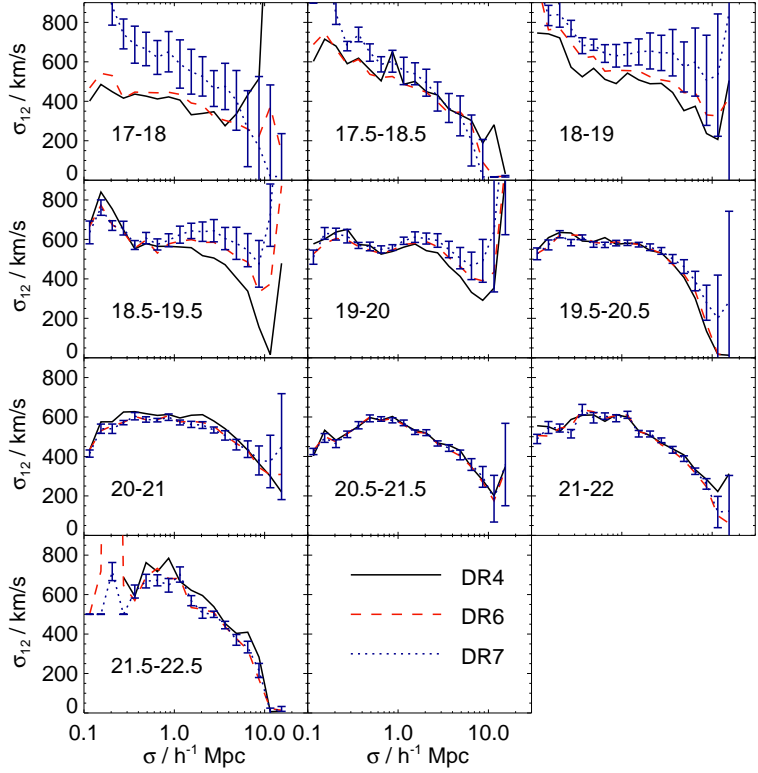}}
\end{center}
\caption{The scale dependence of pairwise velocity dispersions in
the volume-limited samples.} \label{fig:sig12_scale}
\end{figure}

\begin{figure}
\begin{center}
\resizebox{0.75\hsize}{!}{\includegraphics{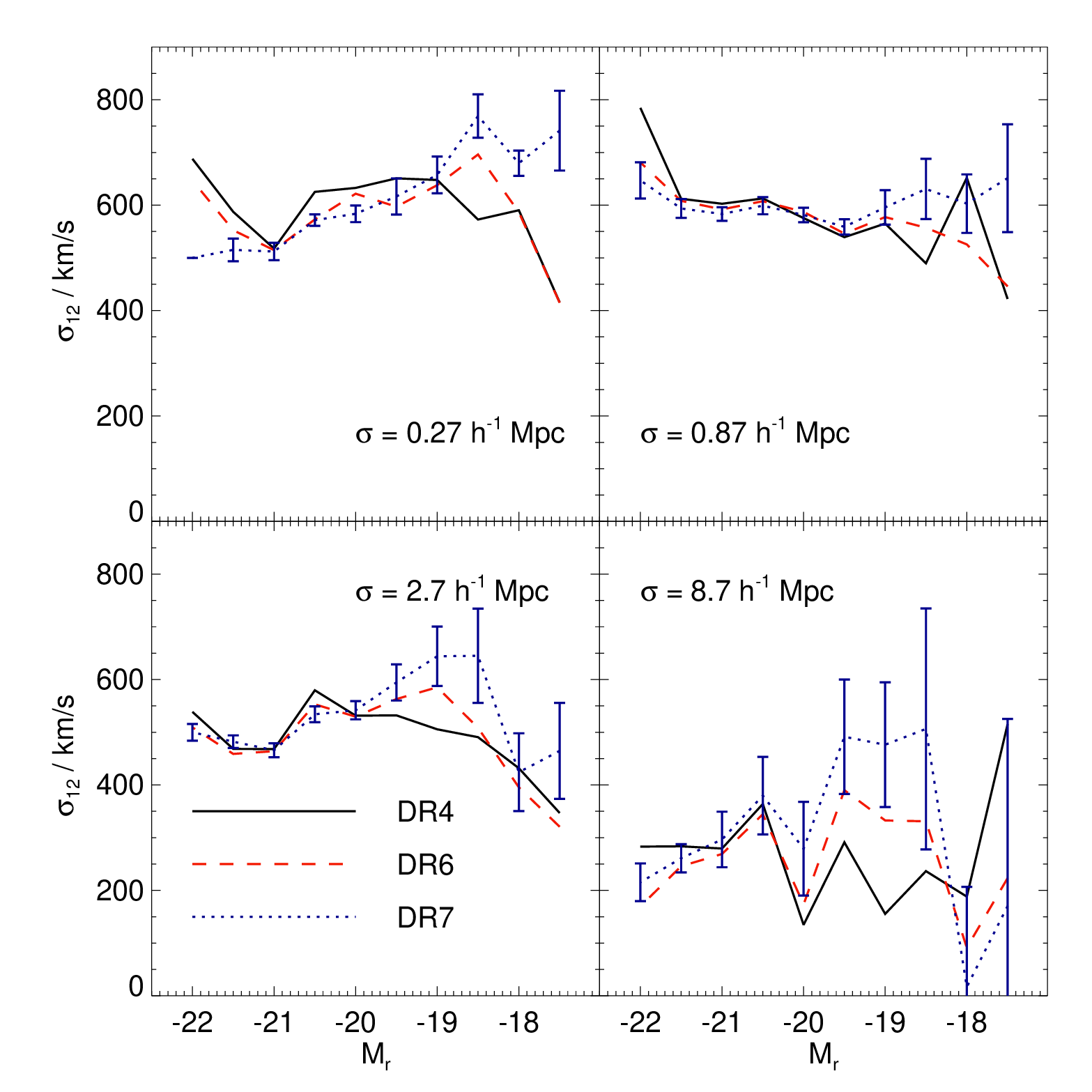}}
\end{center}
\caption{The luminosity dependence of pairwise velocity dispersions
at fixed scales.} \label{fig:sig12_mag}
\end{figure}

$\xi(s)$ is a mixture of real space 2PCF and PVD. The
entanglement can be sorted with the projected 2PCF $w_p$. Measurements of $w_p$ are
shown in Figure~\ref{fig:wp}, actually we cross checked our $w_p$ of DR7 with available
results of \citet{ZehaviEtal2010}, the agreement is excellent except for
the sample VL2 of which our $w_p$ differs at scales $\sigma>\sim 4Mpc/h$.
As seen in Figure~\ref{fig:wp}, obviously that $w_p$ of DR4 and DR6 are in good agreement at
scale range probed in most luminosity bins, $w_p$ of DR6
are slightly larger at large scale around $\sigma\sim 10\hmpc$
in several faint samples but of low significance for the size of error bars.
For VL1 and VL2, their $w_p$ of DR7 are boosted by more than $70\%$ in amplitude relative
to of DR4, but the shape does not change. For subsamples in other luminosity bins,
their $w_p$ are stable against data version, though for VL1+ and VL2+ there are
some minor changes within errorbars.

Figure~\ref{fig:sig12_scale} demonstrates
the scale dependence of PVDs $\sigma_{12}$ of different luminosity samples while
Figure~\ref{fig:sig12_mag} is of the luminosity dependence of PVDs
measured at scales of $\sigma=0.27, 0.87, 2.7, 8.7\hmpc$ respectively.
$\sigma_{12}$ of subsamples VL1, VL2 and VL2+ of DR7 are significantly different to
measurements of DR4. For VL2+, PVD of DR7 agrees with earlier data at small scales but
then turns to be higher at scales $\sigma>1h^{-1}$Mpc which makes the scale dependence
very weak; for VL2, $\sigma_{12}$ of DR7 roughly keeps the shape of DR4
but has a much larger amplitude. $\sigma_{12}$ of VL1 of DR7 has steeper scale dependence
and stronger amplitude at small scales than results of DR4 and DR6. $\sigma_{12}$
of VL1 subsamples of DR4 and DR6 are rather flat, and do not follow
the general trend that PVDs of galaxy samples with lower luminosities should rise faster
at smaller scales \citep[also see PVDs of SAMs in Fig.5 of][]{LiEtal2007},
but now DR7 reverts VL1 to track.
Comparing distributions in celestial sphere of galaxies in the lowest luminosity bin of
the three SDSS data releases reveals the variation is just induced by
a big structure locates in area roughly of RA $166^\circ-188^\circ$ and
DEC of $16^\circ-26^\circ$ (Figure 1). It is another example of impact of
super structure on clustering analysis of LSS in addition to the Sloan Great Wall.

\citet{LiEtal2007} realised that $w_p$ and PVDs of faint volume-limited samples of
DR4 are too low to match prediction of SAMs. Instructed by the experiment
of \citet{SlosarEtal2006}, they reduced fraction of satellite galaxies in massive halos
in SAMs {\it ad hoc} by around $30\%$ and reproduced approximately the actual
measurements,
which then becomes a serious conflict for people to reconcile between models
and observation.
An eyeball check of our results with the SAMs prediction in \citet{LiEtal2006}
denotes that the amplitude boost in $w_p$ and PVDs of DR7's faint volume-limited samples
roughly compensate for the space between DR4 and SAMs, or at least ameliorate
difficulties in theoretical modelling, although we do not have SAMs data to
quantify the improvement. So unlike the Sloan Great Wall,
the existence of a large structure in the Universe is actually positive
to our working models, which somehow
casts doubts in the proclaimed practice
of cutting off super structures from original data to fit into an unified picture.
After all, it is still early to say which is closer to the true clustering
property of those faint galaxies, we might need a deeper and wider survey than
the present SDSS DR7 to reach good fairness and reduce huge uncertainties.

\section{Summary and Discussion}
By extensive comparison of different data releases of SDSS main galaxy catalogue with
2PCFs in redshift space for flux-limited samples, 2PCFs/monopole of 3PCFs
in redshift space for volume-limited samples, projected 2PCFs and PVDs for
volume-limited samples, we have the following findings about galaxy clustering properties
against the expansion of sky coverage of SDSS.
\begin{enumerate}

\item
2PCF $\xi(s)$ in redshift space of flux-limited sample is extremely robust against
sample volume change, which subsequently secures relevant application; $\xi(s)$
is also insensitive to local structures at low redshift.

\item
2PCFs in redshift space $\xi(s)$ of volume-limited samples of SDSS DR7 in
luminosity bins brighter than
$-M_{r,0.1}=[17,18]$ are in good agreement with earlier data releases at scales
$s< \sim 10\hmpc$. As scale being larger, the consistency is broken for volume-limited samples
fainter than $-M_{r, 0.1}=[19.5,20.5]$, and in general the deviation of DR7 to DR6
and DR4 grows with larger absolute magnitude. Zero crossing points of DR7's
$\xi(s)$ do not differ much to DR6's, but shifts away from DR4's apparently.

\item
Volume-limited samples of SDSS display convergence in $\zeta_0$ at scales
$s<\sim 10\hmpc$ except the one in the faintest luminosity bin; while in the weakly
nonlinear regime, there is no agreement between $\zeta_0$ of
different data releases in all luminosity bins.

\item
Projected 2PCFs $w_p$ of volume-limited samples in luminosity bins brighter than
$-M_{r,0.1}=[18.5,19.5]$ are robust against data version, but for samples in fainter bins,
$w_p$ of DR7 are significantly higher than those of earlier data. Similar phenomenon is
also seen in PVDs, PVDs of the two faintest volume-limited samples also
appear much steeper along scale in DR7 and then become flatter at higher luminosity,
which actually turn to be closer to what SAMs predict shown in \citet{LiEtal2007}.

\item
The faintest volume-limited sample of $-M_{r,0.1}=[17,18]$ is very peculiar, it
suffers of the biggest variance due to enlargement
of sky coverage, agreement of $\xi(s)$ and $\zeta_0$ of DR7 in redshift space
with early data is breached at scales as small as $\sim 3\hmpc$; $w_p$
of the sample is enhanced by around $\sim 70\%$ and PVDs distinguish much in amplitude
and scale dependence from measurements of earlier data.

\end{enumerate}

Fairness of a galaxy sample is assessed by statistical functions, one can
not claim a general fair sample hypothesis without specifying
the statistical measure used. It is possible that a galaxy sample is fair
for one statistical function but not for another function.
With our measurements, we conclude that current SDSS is not able to provide reliable 2PCFs
(both of redshift space and projected) and PVDs of samples of
characteristic luminosity fainter than $L^*$, and third-order
statistics in the weakly nonlinear regime for nearly all volume-limited samples.

For faint volume-limited subsamples, probably due to
their very shallow depths, measurements suffers of greater finite volume effects
such that enlarging sky coverage has larger influence on measurements of
statistics than for bright subsamples.
The inconsistency observed is manifestation of cosmic variance due to
insufficient sample volume. The variances are comparable to the $1\sigma$
jack-knife errorbars which usually are regarded as good
and robust approximation to the true errorbars \citep{ZehaviEtal2002}. Now
it seems that the technique underestimates the true
variance, corresponding results about the habitation of faint
galaxies in halos withdrawn from clustering analysis,
e.g. \citet{LiEtal2007} and \citet{ZehaviEtal2010} are not very concrete.
Conclusions about faint galaxies utilizing galaxy group catalogue constructed from
SDSS DR4 \citep{YangEtal2007, YangEtal2008} might also be problematic, we conjecture
that a new group catalogue from DR7 may provide a very different
paradigm.

In our analysis PVDs are derived under an general assumption that galaxy
pairwise velocities are following closely to exponential distribution. The assumption
might not be exact for satellite galaxies which pairwise velocity distribution
can be better described by Gaussian \citep{Tinker2007}. For galaxies of low luminosity,
they are mostly likely satellite, the obtained $\sigma_{12}$ based on exponential
distribution is biased and so be the relation of PVDs with galaxy luminosity presented
in Figure~\ref{fig:sig12_mag}. Nevertheless, our PVDs of different versions of VL1 are
biased in the same way, the systematical bias will not affect our basic
conclusion that PVD of VL1 of DR7 is very different to what is of DR4.

Recently there are several works applying 3PCF of SDSS
\citep[e.g.][]{SefusattiEtal2006, KulkarniEtal2007, MarinEtal2008,
Marin2010, McBrideEtal2010}, either to help
determining cosmological parameters and galaxy biasing or to diagnose
models of galaxy formation. Some results are using measurements
of volume-limited subsamples of the SDSS main galaxy catalogues in the weakly
nonlinear regime, our analysis however points out that one needs to be
very cautious in taking relevant conclusions.

Another problem worthy of more discussion is the first zero-crossing point of 2PCF.
Of course, part of the problem is induced by finite volume of samples, at least
integral constraint is a serious systematics to low luminosity galaxy subsamples.
But for subsamples with large volume of bright galaxies, absent of anti-correlation
at large scales is still puzzling. Instead of assaulting validity of $\Lambda$CDM models
it is probably better to activate stochastic bias in models of galaxy 2PCF.
Halo model alone can not solve the problem
since at large scales 2PCF in halo model boils down to simple multiplication of
bias parameter with linear 2PCF of dark matter. In the bucket of parameters
of cosmological application, galaxy 2PCF at large scales by default is fully
described by linear bias parameter and 2PCF of dark matter, the single
bias parameter is largely degenerated with some other parameters, such as the
normalization of density fluctuation $\sigma_8$ and the matter density parameter
$\Omega_m$, it is unclear if present estimation of cosmological parameters
is significantly biased by the ignorance of possible exotic
bias \citep[e.g. the proposal of][]{ColesErdogdu2007}.

\begin{acknowledgements}
This work is funded by the NSFC under grants of
Nos.10643002, 10633040, 10621303, 10873035, 10725314 and the Ministry of Science \& Technology
of China through 973 grant of No. 2007CB815402. Authors enjoy
wonderful discussion with Xiaohu Yang, Weipeng Lin and Xi Kang.

This publication also makes use of the {\it Sloan
Digital Sky Survey} (SDSS). Funding for the SDSS and SDSS-II has
been provided by the Alfred P. Sloan Foundation, the Participating
Institutions, the National Science Foundation, the U.S. Department
of Energy, the National Aeronautics and Space Administration, the
Japanese Monbukagakusho, the Max Planck Society, and the Higher
Education Funding Council for England. The SDSS Web Site is http://www.sdss.org/.

The SDSS is managed by the Astrophysical Research Consortium
for the Participating Institutions. The Participating Institutions
are the American Museum of Natural History, Astrophysical Institute
Potsdam, University of Basel, University of Cambridge, Case Western
Reserve University, University of Chicago, Drexel University, Fermilab,
the Institute for Advanced Study, the Japan Participation Group, Johns
Hopkins University, the Joint Institute for Nuclear Astrophysics, the
Kavli Institute for Particle Astrophysics and Cosmology, the Korean
Scientist Group, the Chinese Academy of Sciences (LAMOST), Los Alamos
National Laboratory, the Max-Planck-Institute for Astronomy (MPIA),
the Max-Planck-Institute for Astrophysics (MPA), New Mexico State University,
Ohio State University, University of Pittsburgh, University of Portsmouth,
Princeton University, the United States Naval Observatory, and
the University of Washington.

This publication also made use of NASA's Astrophysics
Data System Bibliographic Services.
\end{acknowledgements}



\end{document}